\documentclass[twocolumn,showpacs, amsmath, amssymb, floatfix,
                nofootinbib,wrapfloat byrevtex]{revtex4}
 \usepackage{graphicx,epstopdf}
 \usepackage[bookmarks=true,%
  bookmarksopen=true,%
  pdfborder={0 0 0},%
  pdfhighlight={/N},%
  linkbordercolor={.5 .5 .5}]{hyperref}

 \newcommand{\beq}[1]{\begin{eqnarray}\label{#1}}
 \newcommand{\eeq}{\end{eqnarray}}

\begin{document}

 \title{An Exact Hairy Black Hole Solution for AdS/CFT Superconductors}

 \author{Ding-fang Zeng}
 \email{dfzeng@bjut.edu.cn}

 \affiliation{Institute of Theoretical physics,
 Beijing University of Technology}

 \begin{abstract}
 We provide an exact hairy black hole solution to an $n+1$ dimensional
 gravity-coupled complex scalar field model. The solution has translationary
 invariant horizon and tunable temperatures. Free energy calculations
 indicate that, there are always temperature ranges
 in which the hairy black hole is thermodynamically stable against
 decaying into its no-hair counterpart. Using this solution
 as an AdS/CFT superconductor model, we get potentially useful
 critical temperature v.s. dimensions of order parameter operator
 and conductivity-frequency relations typical of other similar models.
 \end{abstract}


 \date{\today}

 \maketitle

 %\tableofcontents

 Studies in applied holographic theories, such as finite
 temperature AdS/QCD models' building \cite{thermoQCD-KSS0405,thermoQCD-GKN0804,thermoQCD-GNPR0804}, AdS/CFT
 superconductor's description \cite{adscftSuperconduct-Gubser,adscftSuperconduct-HHH}, or other quantum
 phase transitions' exploration \cite{quantumCriticalTrans-HKSS0701,quantumCriticalTrans-HK0704,quantumCriticalTrans-HH0706} all call for solutions to the scalar field model coupled with gravity,
 \beq{}
 S=\frac{1}{16\pi G_N}\int{dx^{n+1}}\sqrt{-g}(R
 -\partial_\mu\Psi\overline{\partial^\mu\Psi}-V[\Psi])
 \label{gravitonDilatonAction}
 \eeq
 of the form
 \beq{}
 ds^2=e^{2A}(-hdt^2+d\vec{x}\cdot{d\vec{x}})+h^{-1}du^2\,,\;\Psi=\Psi(u)
 \label{metricAnsatz}\\
 u\in(0,\infty),\;A\rightarrow u\,,\;h\rightarrow1\,\mathrm{as}\,u\rightarrow\infty
 \\
 \exists u_0\geqslant0\,,\;h(u_0)=0\,,\;h'(u_0)\mathrm{\;is\;finite}
 \eeq
 However, to this day people still have no exact
 analytical solutions of the desired form, even for
 the simplest potential $V=m^2|\Psi|^2$. As we know, the best known
 solutions in this area are \cite{ChamblinReall99} and
 \cite{MTZblackhole}. The former can be derived out from
 higher dimensional AdS-Schwarzschild solutions through
 compactification \cite{GubserNellore0804}. At the desired
 dimension its asymptotic is not AdS type. The latter's
 horizon has special topologies and the asymptotic is only
 locally AdS type \cite{adscftSuperconduct-KPS0902}. Few days
 before this work is finished, we note \cite{thermoQCD-soundspeed-jetquencch0804}
 in the arxiv, which provides finite temperature solutions
 to a softwall AdS/QCD model. But its temperature cannot be tuned
 freely for fixed scalar potentials.

 Difficulties to find analytical solutions in this model is due to
 the non-linearity of the equations of motion, which follows by minimizing the action \eqref{gravitonDilatonAction}
 \beq{}
 {tt,xx}&:&h''+nh'A'=0\label{EinsteinEq1}\\
 tt,uu&:&A''+\frac{1}{n-1}\Psi'^2=0\label{psipsEq}\\
 \mathrm{Eins.Eq.}{uu}&:&(n-1)\big[h'A'+nhA'^2\big]=h\Psi'^2-V\label{Veq}\\
 \mathrm{scalar\;eom}&:&2(\Psi'he^{nA})'e^{-nA}=\frac{dV}{d\Psi^\star}.
 \eeq
 Note we will focus on constant phase solutions since any non-trivial
 phase profile will cause the action deviate from minimal configuration.
 We checked that the scalar field equation of motion can be derived
 out from the three components of the Einstein equation.
 So, only the former three of the above four equations are independent.
 Of them, the first one can be integrated once to give
 \beq{}
 A=-\frac{1}{n}\ln h'+\mathrm{const.}\label{Aexpression}
 \eeq
 Usually given forms of the potential function,
 even of the simplest quadratic type, to integrate
 equations \eqref{EinsteinEq1}-\eqref{Veq} is almost impossible. However, if we
 ask the question from an inverse direction ---
 given function $h$ of desired asymptotics, can we
 find out the potential function's form explicitly?
 --- we may have different gains.

 In most cases, when we write down expressions for $h$ with
 some asymptotics, substitute into eq\eqref{Aexpression}
 and get $A$, into eq\eqref{psipsEq} and get $\psi'$,
 integrate and get $\psi(u)$, combining these things
 into eq\eqref{Veq} and find $V(u)$, we could only
 find $V(u)-\Psi(u)$ correspondence but not
 the explicit form of $V[\Psi]$. However, after some trial and errors, we find
 that the following metric functions and potential
 can be worked out simultaneously
 \beq{}
 &&\hspace{-5mm}h[u]=1-e^{-n(u-u_0)/\ell}\frac{1+n/(n+k)e^{-k(u-u_0)/\ell}}{1+n/(n+k)}
 \label{exactHfunc}
 \\
 &&\hspace{-5mm}A[u]=\frac{u}{\ell}-\frac{1}{n}\ln\big[1+e^{-k(u-u_0)/\ell}\big]
 \label{exactAfunc}\\
 &&\hspace{-5mm}\Psi[u]=\Big(\frac{n-1}{n/4}\Big)^\frac{1}{2}\arctan[e^{-k(u-u_0)/2\ell}]
 \label{exactPsifunc}\\
 &&\hspace{-5mm}V[\Psi]=-\frac{(n-1)k}{\ell^2}\Big[\frac{n}{k}+(2-\frac{k}{n})\sin^2[\hat\psi]+\frac{2k}{n}\sin^4[\hat\psi]
 \nonumber\\
 &&\rule{9mm}{0pt}+\frac{k^2}{n(2n+k)}(\sin^2\hat\psi-2\sin^4\hat\psi)\tan^\frac{2n}{k}\hat\psi\Big]
 \label{scalarPotential}\\
 &&\hspace{-5mm}\hat\psi=\Big(\frac{n/4}{n-1}\Big)^\frac{1}{2}|\Psi|
 \eeq
 This means that we find exact hairy black hole
 solutions to the complex scalar field model
 featured by the potential \eqref{scalarPotential}.
 Obviously, these solutions have tunable temperature and the same singularity structure as the
 usual AdS-Schwarzschild black holes with spatial-flat horizons. Although
 the potential function here is not so easily looking, the fact
 that it is periodical function of the main argument $\hat{\psi}$
 makes us believe that,
 the probability of finding it in the landscape of M-theory superconduction \cite{superconductingLandscape0901}
 would not be too small. We know, in string/M theory
 compactifications, there are many moduli fields which
 are constrained to take values periodically, e.g $(0,2\pi)$.
\begin{figure}[h]
\includegraphics[clip=true,bb=63 690 306 790]{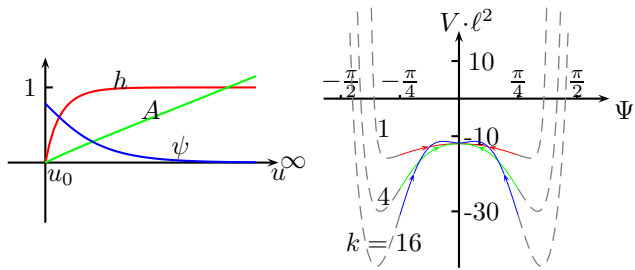}
\caption{Metric function, scalar field profile and
potentials of the latter, $n=4$ as examples. Other dimensions are similar.
For the potentials, we let $\Psi$ take real values. When it takes complex values,
the potential can be obtained by rotating
this figure around the vertical axes. The dashed part of
the curve corresponds to field values possible only inside the
black hole horizon.
}
\label{figScalarPotential}
\end{figure}

For small $k$, our potential has similar features as the standard
model Higgs field does. The difference is, in the standard
model, the Higgs field will uniformly sit at the global minimal
of the potential; while
the scalar field here has non-trivial spatial
profile. From the black hole horizon to infinite,
$\Psi$ field tries its best to climb up the local maximal of the potential,
see FIG \ref{figScalarPotential}. This constitutes a non-trivial hair
of the black hole.  Asymptotically,
$\Psi\rightarrow0$, $V[\Psi]\approx-n(n-1)\ell^{-2}+m_{eff}^2|\Psi|^2$,
with
\beq{}
m_{eff}^2=
-(2-\frac{k}{n})(n-1)k\ell^{-2}
\label{massEffective}
\eeq
So near the AdS boundary, our model has little difference from
the simple gravity-coupled scalar
field theory with quadratic potentials. When
$k<2n$, the effective mass parameter (mass square) of
the scalar is negative. But if at the same time
\beq{}
n-n\sqrt{1-\frac{n/4}{n-1}}<k<n+n\sqrt{1-\frac{n/4}{n-1}}
\eeq
this negative mass square will not lead to instabilities.
Because in this case, the Breitenlohner-Freedman bound $m_{eff}^2\ell^2>-n^2/4$
is always satisfied. When $k>2n$, the effective mass square is positive.
So for large $k$, $\Psi=0$ is a meta-stable point of the potential, see also FIG \ref{figScalarPotential}.

Since the hair of the black hole breaks symmetry of the model
associated with the phase rotation of the scalar field, coupling
this scalar field to some $U(1)$ gauge field will give us an ideal
model of AdS/CFT superconductors. But before turning
to that topic, let us examine thermodynamics of this
solution further. Following reference
\cite{BalasubramanianKraus99}, we calculate
the entropy, temperature and energy of the solutions as follows
\beq{}
S&=&\frac{V_{n^{\!_-}}e^{(n-1)A}}{4G_N}\Big|_{u_0}=\frac{V_{n^{\!_-}}e^{(n-1)u_0/\ell}}{4G_N}\;,\;n^-=n-1
\\
T&=&\frac{e^Ah'}{4\pi}\Big|_{u_0}=\frac{n}{2\pi}\frac{n+k}{2n+k}e^{\frac{u_0}{\ell}}\cdot\ell^{-1}\\
E&=&-\frac{V_{n^{\!_-}}n^{\!_-}}{8\pi G_N\ell}
\Big\{\big[e^{nA}hA'\big]^\infty_{\stackrel{hairy}{AdS.bh}}-\big[e^{nA}hA'\big]^\infty_{pure-AdS}\Big\}
\nonumber\\
&=&\frac{V_{n^{\!_-}}n^{\!_-}\cdot\ell^{-1}}{8\pi G_N}\Big[\frac{n+k}{2n+k}e^{\frac{nu_0}{\ell}}-\frac{k}{n}e^{\frac{ku_0}{\ell}}e^{\frac{(n\!-\!k)u}{\ell}}_{\;\;\;u\rightarrow\infty}\Big]
\label{energyHoloRenorm}
\eeq
By adding boundary term and counter
term to the action \eqref{gravitonDilatonAction},
and make strict holographic renormalization
treatment, we will get similar results.
It can be checked that, $[dE\neq TdS]_{\mathrm{vary}.\;u_0}$.
This is not a catastrophic. For example, in reference\cite{BuchelAndLiu0305}
the same question happens.
In that reference, violation of the first law
is attributed to some chemical potential associated with
the scalar particle numbers, i.e. $[dE=TdS+\mu dN]_{\mathrm{vary}.\;u_0}$.
In the current paper, we think the same expanation helps.

\begin{figure}[h]
\includegraphics[clip=true,bb=60 666 368 790,scale=0.75]{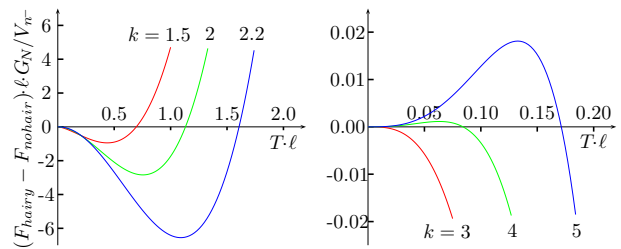}
\caption{Free energy difference between the
hairy AdS-black hole and that of the no-hair partners for $n=3$
case. For other values of $n$, the results are similar.}
\label{figFreeEnergy}
\end{figure}
The energy definitions \eqref{energyHoloRenorm} involve regularizations
in which an infinite part was subtracted. However, for $k<n$
models, this regularization may subtract too much so that
the resulting energy becomes negative. For $k>n$ there are
contrary questions. Tuning the subtractions
appropriately (assure positivity of the energy but hold
the $k$ relevant part in the finite part),
we will get
\beq{}
E&=&\frac{V_{n^{\!_-}}n^{\!_-}\cdot\ell^{-1}}{8\pi G_N}
\Big[\frac{n+k}{2n+k}e^{\frac{nu_0}{\ell}}-\frac{k}{n}e^{\frac{ku_0}{\ell}}\Big]
\eeq
As results, the free energy $F\equiv E-TS$ of system reads
\beq{}
F&=&-\frac{V_{n^{\!_-}}\cdot\ell^{-1}}{8\pi G_N}
\Big[\frac{n+k}{2n+k}e^{\frac{nu_0}{\ell}}+\frac{k n^{\!_-}}{n}e^{\frac{k u_0}{\ell}}
\Big]\label{freeEnergyAnaExpression}
\eeq
Comparing this result with that of the no-hair counterpart,
we see that, for $k<n$ case, referring to the left part of figure \ref{figFreeEnergy},
the hairy black hole at lower temperatures has
lower free energy than their no-hair partners so
the former is favored thermodynamically at low temperatures;
for $k>n$ case, referring to the right part the same
figure, the hairy black hole at higher temperatures
have lower free energy so is favored thermodynamically
at high temperatures.

\begin{figure}[h]
\includegraphics[clip=true,bb=64 687 368 790,scale=0.75]{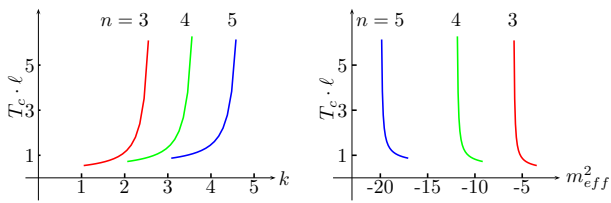}
\caption{Effects of $k$ and/or the effective mass
parameters on the phase transition
critical temperatures for different space time dimensions.}
\label{figTemperatureCritic}
\end{figure}

For $k<n$ models, if at high temperatures
the system lies at the no-hair phase in which it is symmetric
with respect to the $U(1)$ phase rotation of the scalar field,
as temperature lowers, it will undergo a phase transition
and break the symmetry spontaneously, i.e. the scalar field
attains non-trivial profile with specific phases. The phase
transition temperature depends on $n$ and $k$. Figure
\ref{figTemperatureCritic} displays
this dependence explicitly, from which we easily see that
increasing $|m_{eff}|$ will increase the phase
transition temperature. When translated into
the dual field theories, this may be a potentially useful result
for looking for superconductors with
higher critical temperatures. Of course it should be noted
that, $m_{eff}$ in AdS/CFT models corresponds to the scaling
dimension of order parameter operators in the Ginzburg-Landau superconductor models.
For $k>n$ models, from the right part of
figure \ref{figFreeEnergy} we see that, the symmetry breaking
(hairy) phase has lower free energies so is favored at
higher temperatures. This is contrary to the actual
case of the real superconductors. So in the
following explorations, we will neglect this case.

To build superconductor models in holographic
languages, we couple our scalar field to some $U(1)$ gauge field
\beq{}
S=\frac{1}{16\pi G_N}\int{dx^{n+1}}\sqrt{-g}(R
-\frac{1}{4}F_{\mu\nu}^2
\nonumber\\
-(\partial_\mu-iqA_\mu)\Psi(\partial^\mu+iqA^\mu)\Psi^\star-V[\Psi])
\label{fullAction}
\eeq
By previous analysis, we know
the scalar field in $k<n$ models will condensate and break the local $U(1)$ symmetry of
the system spontaneously at low temperatures. As
results, materials described by the dual conformal (approximately) field theory
will go into superconduction phase. At this phase,
no direct current is allowed in the material. But we can use some small
electro-magnetic field to perturb the material and measure
its response. By AdS/CFT correspondence, this means that we
perturb the background geometry \eqref{exactHfunc}-\eqref{exactPsifunc}
by small, but uniform gauge field $A_mdx^m$
\beq{}
A=e^{-i\omega t}a_x(u)dx\,,\;
\eeq
Note we do not include backreaction
of the background geometry. For this perturbation,
the linearized Maxwell equation reads
\beq{}
a_x''+a_x'\big[(n-2)A'+\frac{h'}{h}\big]
+a_x\big[\frac{\omega^2}{h^{2}e^{2A}}-\frac{2q^2|\Psi|^2}{h}\big]=0
\label{emPerturbEq}
\eeq
Near horizon of the black hole, causality requires
\cite{fallingBC} the perturbing field $a_x$ satisfy
falling boundary condition,
\beq{}
a_x\propto e^{-i\omega/(h_0'e^{A_0})\ln(u-u_0)}
\label{nearHorizonBC}.
\eeq
While in the asymptotically infinite region,
from the background field expressions \eqref{exactHfunc}-\eqref{exactPsifunc}
and the perturbation eq\eqref{emPerturbEq} we can show that
\beq{}
a_x=a_{x(0)}+a_{x(1)}e^{-(n-2)u}+\cdots
\label{nearAsympInftyBC}.
\eeq
According to the AdS/CFT dictionary, in the dual CFT side,
$a_{x(0)}$ and $a_{x(1)}$ correspond to external field strength
and the resulting current respectively,
so the conductivity there can be calculated
through
\beq{}
\sigma=\frac{j_x}{E_x}=-\frac{i}{\omega}\frac{a_{x(1)}}{a_{x(0)}}
\eeq

In the normal phase $\Psi\equiv0$, eq\eqref{emPerturbEq} can be
changed into
\beq{}
\frac{d^2}{du^2_\star}a_x+\omega^2e^{(2n-6)A}a_x=0
\;,\;
du_\star=du/[he^{(n-2)A}]
\label{nmphasePEqsimplified}
\eeq
where $h$ and $A$ should be given by letting
$k=0$ in eqs\eqref{exactHfunc} and \eqref{exactAfunc}.
For $n=3$, this equation can be solved exactly
\beq{}
a_x=e^{-i\omega u_\star}
\;,\;
u_\star=\int\frac{-de^{-(u-u_0)}}{1-e^{-3(u-u_0)}}
\label{axD3exactsol}
\eeq
This solution is selected from two possibilities
by the falling boundary condition \eqref{nearHorizonBC}.
Expanding it into the form of \eqref{nearAsympInftyBC}
around $u\rightarrow\infty$, we find that for $n=3$ case,
the normal phase conductivity reads
\beq{}
\sigma(\omega)=1
\eeq
For $n\geqslant4$, we do not know how to solve
eq\eqref{nmphasePEqsimplified} analytically. But numerics tell
us that $\sigma(\omega)\propto\omega^{n-3}$ as $\omega\rightarrow\infty$.

\begin{figure}[h]
\includegraphics[clip=true,bb=64 671 368 790,scale=0.75]{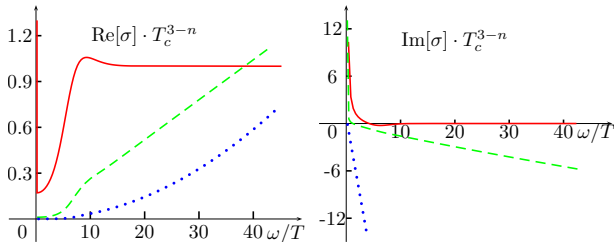}
\caption{Conductivity-frequency dependence for $n=3$(red), $4$ (green)
and $5$ (blue) models at each own's superconduction
critical temperature $T=T_c$. All three curves has $k=2$,
$q=1\ell^{-1}$. The conductivity of $n=4$ model has been
scaled a factor of $1/30$, while that of the $n=5$ models
is scaled a factor of $1/900$.
}
\label{figsig-f-dimension}
\end{figure}

In the superconduction phase, $\Psi\neq0$. In this phase even for
$n=3$, eq\eqref{emPerturbEq} cannot be solved analytically.
But we can solve it numerically and get the
conductivity approximately as
\footnote{We thank very much to Dr/Professor C. P. Herzog
to provide us his notebook on numerical details of the
work \cite{adscftSuperconduct-HHH}, from which we learned this technique.}
\beq{}
\sigma=\frac{i}{\omega}\frac{a'_x}{a_x}\frac{e^{(n-2)u}}{n-2}\Big|_{u\rightarrow\infty}
\eeq
Figure \ref{figsig-f-dimension} displays our numerical results
of conductivities for some specific $n$, $k$ and $q$ values.
Two features of the figure should be noted here. The first is,
there is a $\delta$ function peak in the real part of
the conductivities as the result of Kramers-Kronig relation
\beq{}
\mathrm{Im}[\sigma(\omega)]=-\frac{1}{\pi}\int_{-\infty}^\infty
\frac{\mathrm{Re}[\sigma(\tilde\omega)]}{\tilde\omega-\omega}d\tilde\omega
\eeq
and the fact that imaginary part of the conductivity has a pole
of the form Im$[\sigma(\sigma)]\sim\frac{1}{\omega}$. This is just
the usual definition of superconduction, i.e. infinite DC
conductivity. Note this infinity cannot be attributed to
translation invariance, since in the calculation we fixed
the background geometry. As results, the translation invariance
is broken implicitly, see reference \cite{adscftSuperconduct-HHH}.
The second is that, conductivities of different dimensional
models behave hierachically, $\sigma_3\ll\sigma_4/T_c\ll\sigma_5/T_c^2$.
Note to plot the conductivity curves of different dimensional models
in the same figure and make them look clear enough, we multiply
different numerical factors on the conductivities.
On the first feature, our $n=3$ result coincides with those based
on AdS abellian-higgs models, such as \cite{adscftSuperconduct-HHH,adscftSuperconduct-RobertsHartnoll0805,
adscftSuperconduct-HorowitzRoberts0810,adscftSuperconduct-GubserNellore0810,
adscftSuperconduct-GubserPufu0811}; our $n=4$ result is similar to
those based on D3/D7 brane models\cite{braneSuperconduct-KernerBasu},
see also reference \cite{adscftSuperconduct-HorowitzRoberts0810}.
On the second feature, our results for $4$ dimensional models
are similar to that of reference \cite{adscftSuperconduct-HorowitzRoberts0810}
so can be supported by that paper.

A little summary: we construct exact analytical solutions
to an $n+1$ dimensional gravity-coupled complex scalar field
model. When the scalar field in this model is coupled to
$U(1)$ gauge field, the resulting system provide holographic
descriptions for superconductors. We calculate conductivities
in this model and find results typical of other holographic
superconductors. But the superconduction mechanism
in our model is not totally the same as that proposed by
Gubser \cite{adscftSuperconduct-Gubser}, in which the scalar field
condensates due to it's coupling with the gauge field. Our
scalar field condensates due to its self-interaction,
the potentials of which have the typical Mexican hat shape.
It is worth to note that existence of hairy black
hole solutions as we provide in this paper does not violate
the no hair theorem \cite{nohairTheorem-hertog} refined by
Hertogs. The main reason is that our solutions
do not satisfy the positive energy condition. Few months
after the first version of this paper, reference \cite{moreHairyBHexample} and
\cite{nohairTthorem-JB0906} appears on arxiv. \cite{moreHairyBHexample}
used method similar to us and construct more hairy black hole
solutions. While \cite{nohairTthorem-JB0906} proposed a new
version of no hair theorem in which the existence of
our solutions can be understood naturally.

 \end{document}